\documentclass[]{jkas} 

\def\year{2024} 
\def\volume{??} 
\def\issue{??} 
\def\beginpage{1} 
\def\received{September 23, 2024} 
\def\accepted{December 20, 2024} 
\def\published{?? ??, 2024} 
\date{Received \received; Accepted \accepted; Published \published}

\title{%
Understanding the Radial Acceleration Relation of Dwarf Galaxies with Emergent Gravity
}

\author[1]{Sanghyeon Han}{}
\author[1,2,3$\star$]{Ho Seong Hwang}{0000-0003-3428-7612}
\author[4]{Youngsub Yoon}{0000-0002-0096-4702}

\affil[1]{Department of Physics and Astronomy, Seoul National University, Gwanak-gu, Seoul 08826, Republic of Korea}
\affil[2]{SNU Astronomy Research Center, Seoul National University, 
Seoul 08826, Republic of Korea}
\affil[3]{Australian Astronomical Optics - Macquarie University, 105 Delhi Road, North Ryde, NSW 2113, Australia}
\affil[4]{Department of Physics and Astronomy, Sejong University,
209 Neungdong-ro Gwangjin-gu, Seoul 05006, Republic of Korea}

\def\corrauthor{%
H.S. Hwang, \email{hhwang@astro.snu.ac.kr}
}

\def\runningauthor{%
Han et al.
}

\def\runningtitle{%
RAR of dwarf galaxies with emergent gravity
}

\def\keywords{%
astronomy --- astrophysics --- gravitation --- galaxies: dwarf --- galaxies: kinematics and dynamics
}

\def\abstracttext{%
We examine whether the radial acceleration relation (RAR) of dwarf galaxies can be explained by Verlinde's emergent gravity. This is the extension of \citet{2023CQGra..40bLT01Y}, which examines the RAR of typical spiral galaxies, to less massive systems. To do this, we compile the line-of-sight velocity dispersion profiles of 30 dwarf galaxies in the Local Group from the literature. We then calculate the expected gravitational acceleration from the stellar component in the framework of the emergent gravity, and compare it with that from observations. The calculated acceleration with the emergent gravity under the assumption of a quasi-de Sitter universe agrees with the observed one within the uncertainty. Our results suggest that the emergent gravity can explain the kinematics of galaxies without introducing dark matter, even for less massive galaxies where dark matter is expected to dominate. This sharply contrasts with MOND, where a new interpolating function has to be introduced for dwarf galaxies to explain their kinematics without dark matter.
}

\begin{document}
\jkashead 


\section{INTRODUCTION\label{sec:intro}}

Understanding the dynamics of galaxy is crucial in the study of structure formation and evolution in the universe. Specifically, the difference in galaxy rotation curves between observations and predictions from baryonic matter suggests the existence of dark matter surrounding galaxies \citep{1983SciAm.248f..96R,1983Sci...220.1339R}; see also \citet{1937ApJ....86..217Z} for the proposition of dark matter. Other observations, such as cosmic microwave background radiation \citep{2016A&A...594A..13P} and gravitational lensing of galaxies and clusters \citep{2015MNRAS.449..685H}, also support the idea of dark matter. 
This idea even leads to the existence of galaxy-like structures made up entirely of dark matter (e.g. dark galaxies in \citealt{Minchin05} and \citealt{LeeG24}). 
Despite the importance of dark matter in understanding the structure formation and evolution in the universe, the physical origin and properties of this invisible matter remain elusive \citep{2016PDU....12...56B,2022NewAR..9501659P}. 

One scientific pursuit aimed at comprehending the enigma of dark matter is to modify the gravitational theory, such as the Modified Newtonian Dynamics (MOND) proposed by Milgrom in the early 1980s \citep{1983ApJ...270..365M,1983ApJ...270..371M,1983ApJ...270..384M}.  This proposition suggests discarding the existence of dark matter. Although MOND has been successful in fitting the rotation curves of galaxies \citep{2016PhRvL.117t1101M}, it still lacks clarity regarding the physical mechanism underlying it. This is reflected by the introduction of `interpolating function', which is not derived from a physical theory, but arbitrarily put by hand to fit the rotation curves.

We shift our focus to "emergent gravity", a theory proposed by Verlinde as an alternative perspective on gravity \citep{10.21468/SciPostPhys.2.3.016}. According to Verlinde, the observed phenomena attributed to dark matter are not the result of any physical matter but arise due to the emergent nature of gravity and spacetime, which leads to the volume contribution to entanglement entropy. This emergent nature gives rise to the "apparent" dark matter content, rather than an actual substance adopted in the standard cosmological model, known as "$\Lambda$ Cold Dark Matter" ($\Lambda$CDM) model.  However, the properties of our universe may not be fully described by the $\Lambda$CDM cosmological model, which requires further independent evaluation of the model (e.g. \citealt{2023ApJ...953...98D}).

The emergent gravity has been studied in the context of several phenomena such as the dynamics of early-type galaxies \citep{2018MNRAS.473.2324T}, weak gravitational lensing \citep{2017MNRAS.466.2547B}, the radial acceleration relation for disk galaxies \citep{2023CQGra..40bLT01Y}, and the internal dynamics of dwarf spheroidal galaxies (dwarf spheroidal galaxies,  \citealt{2018MNRAS.477.1285D}). An important feature of the emergent gravity is that it provides a natural physical basis for understanding galaxy rotation curves without inserting an arbitrary interpolating function, unlike MOND. 

Dwarf galaxies provide a promising way to test cosmological models because of the large difference between the dynamical mass and the baryonic mass (e.g. \citealp{2011AJ....141..193O}). The observed difference indicates the presence of unseen mass, typically attributed to dark matter in the current models \citep{2022NatAs...6..659B}. In particular, dwarf galaxies are particularly sensitive to the effect of dark matter in both the inner and outer regions \citep{2022NatAs...6...35L}. Furthermore, dynamical studies of dwarf galaxies could be extended to even lower masses, posing significant challenges to the standard cosmological model \citep{2018MNRAS.477.1285D,2022NatAs...6..897S}. Thus, dwarf galaxies provide a unique opportunity to test alternative cosmological models that do not rely on dark matter.

This study aims to predict the gravitational acceleration for dwarf spheroidal galaxies in both the Newtonian and Verlinde frameworks, and to compare them with the observed one. Section 2 outlines the data and methodology used in this work. Section 3 represents the results of our study and analysis. In Sections 4 and 5, we discuss our results and draw conclusions, respectively.

\section{DATA AND METHOD\label{sec:datamethod}}

Here we describe the data we use, and explain the assumptions and calculations for gravitational acceleration in the Newtonian and Verlinde frameworks.

\subsection{PROPERTIES AND STELLAR-KINEMATIC DATA OF DWARF SPHEROIDAL GALAXIES\label{sec:properties}}

To compute the gravitational acceleration in dwarf spheroidal galaxies, we need to obtain the radial profiles of their dynamical masses. We use the radial profiles of velocity dispersions to derive the mass profiles. To do this, we compile the data from the literature. First, \citet{2009ApJ...704.1274W} provide the radial profiles for the line-of-sight velocity dispersion for eight classical dwarf spheroidal galaxies. Dwarf galaxies in the Local Group have generally low gas content because of ram pressure stripping \citep{2012MNRAS.420.1714S,2019A&A...624A..11H,2021ApJ...913...53P}. This means that most baryonic content in these galaxies is in the stellar component. 

We also add the line-of-sight velocity dispersion profiles for Andromeda II \citep{2012ApJ...758..124H}, NGC 205 \citep{2006AJ....131..332G}, NGC 147, and NGC 185 \citep{2010ApJ...711..361G}. We find that the rotational velocity of these galaxies is not negligible for the contribution to their dynamics, except for Andromeda II (see Figure 6 and Table 3 in \citet{2006AJ....131..332G}; Figure 3, Tables 4 and 5 in \citet{2010ApJ...711..361G}; Figure 6 and Table 3 in \citet{2012ApJ...758..124H}). 

\begin{figure*}[t]
\centering
\includegraphics[width=150mm]{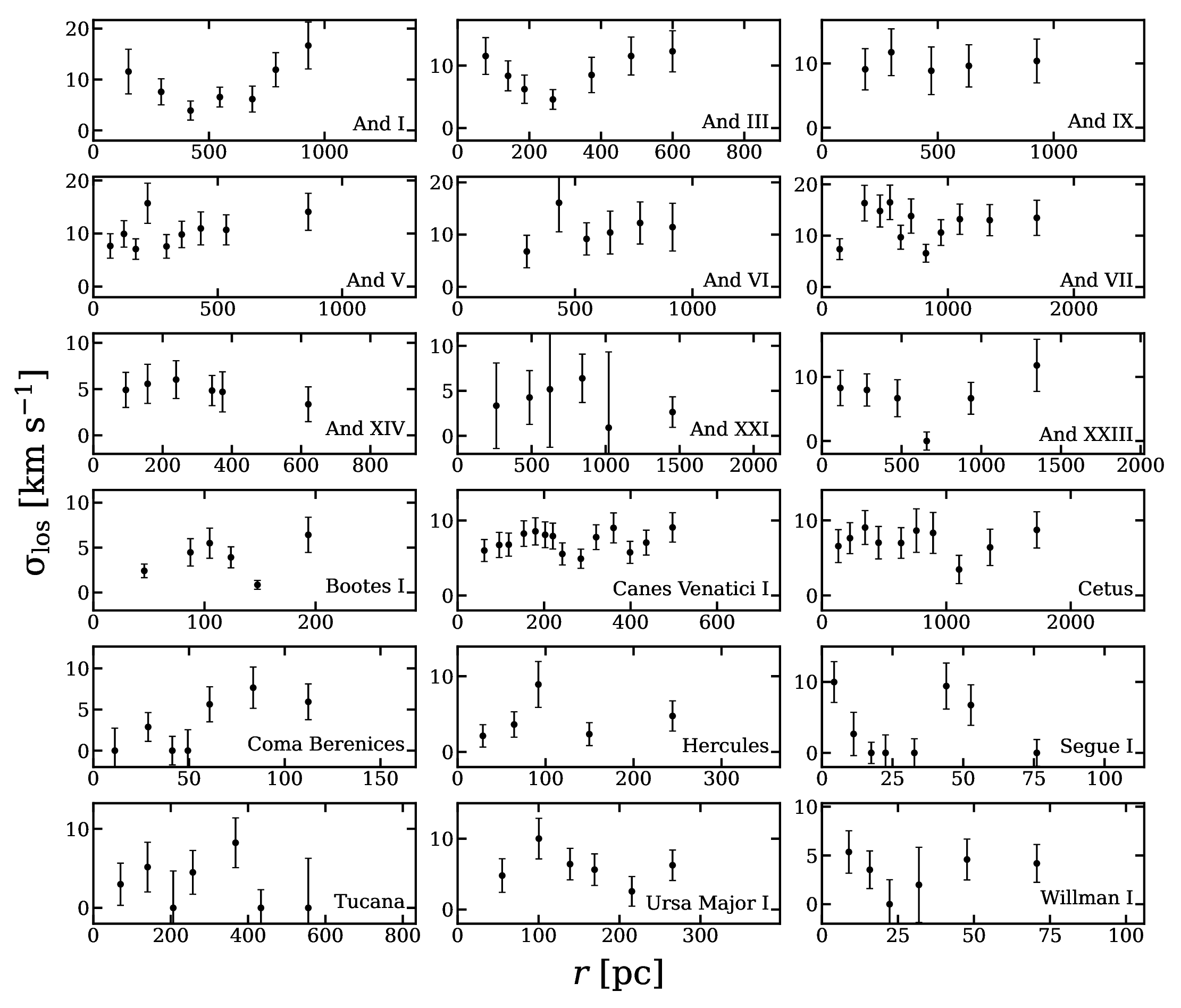}
\caption{Radial profiles of the velocity dispersion are estimated by maximum likelihood estimation adopting \citet{2006AJ....131.2114W}. The estimated velocity dispersion is often recorded as zero in some bins where the true dispersion remains unresolved due to limitations in the number of available data \citep{2015ApJ...801...74G}.\label{fig:profiles}}
\vspace{5mm} 
\end{figure*}

To increase the sample size, we also derive the velocity dispersion profiles from the velocity estimates in the literature. Among 63 dwarf spheroidal galaxies used in \citet{2017ApJ...836..152L}, 41 galaxies have spectroscopic observations except 12 galaxies mentioned above. We apply the maximum likelihood estimation described in \citet{2006AJ....131.2114W} to these data, and obtain the velocity dispersion profiles. We set the number of radial bins to be approximately the square root of the number of member stars of each dwarf spheroidal galaxies, $N_{\rm{bin}} \sim \sqrt{N_{\star}}$. Because of small number statistics for some dwarf spheroidal galaxies, we consider only the galaxies with the number of radial bins greater than five for reliable measurements of radial profiles. We impose the additional condition that the total number of stars with velocity estimates should be larger than 30 to avoid the small number statistics. In the result, 18 galaxies are added to our analysis. The physical properties of these 30 galaxies including distance, half-light radius and V-band luminosity reported in \citet{2017ApJ...836..152L}, are listed in Table \ref{tab:table1}. Figure \ref{fig:profiles} shows the line-of-sight velocity dispersion profiles we derive for 18 dwarf spheroidal galaxies. Table~\ref{tab:table2} provides the galactocentric radius, stellar velocity dispersions, and dispersion uncertainties. It should be noted that we exclude the data points with a velocity dispersion of 0 km s$^{-1}$, which are introduced because of observational uncertainty.

\begin{table*}
\caption{Properties and stellar-kinematic data of dwarf spheroidal galaxies\label{tab:table1}}
\centering
\renewcommand{\arraystretch}{1.32}
\begin{tabular}{lccccccc}
\toprule
Galaxy        & Distance & $\rm{log}$$({L_{\rm{V}}})$ & $r_{\rm 1/2}$ & $N_{\star}$ & Reference for & Reference for  & Notes\\
              &   (kpc)  &        ($L_\odot$)         &     (pc)      & & velocity data & $\sigma_{\rm{los}}$ profile & \\
\midrule
Carina     & 105$\pm 6$    & 5.57$\pm 0.20$ & 273$\pm 45$ &  774 & (2) & (3) & \\

Draco      & 76$\pm 5$     & 5.45$\pm 0.08$ & 244$\pm  9$ &  413 & (4) & (3) & \\

Fornax     & 147$\pm 12$   & 7.31$\pm 0.12$ & 792$\pm 58$ & 2483 & (2) & (3) & \\

Leo I  & 254$^{+19}_{-16}$ & 6.74$\pm 0.12$ & 298$\pm 29$ &  328 & (5) & (3) & \\

Leo II     & 233$\pm 15$   & 5.87$\pm 0.12$ & 219$\pm 52$ &  171 & (6) & (3) & \\

Sculptor   & 86$\pm 6$     & 6.36$\pm 0.20$ & 311$\pm 46$ & 1365 & (2) & (3) & \\

Sextans    & 86$\pm 4$     & 5.65$\pm 0.20$ & 748$\pm 66$ &  441 & (2) & (3) & \\

Ursa Minor & 76$\pm 4$     & 5.45$\pm 0.20$ & 398$\pm 44$ &  212 & (7) & (3) & \makecell{tidally disrupted (8)$^\dagger$ \\ high ellipticity}\\

Boötes I         &  66$\pm 3$  & 4.29$\pm 0.08$ & 283$\pm  7$ &   37 & (9) & (1), (11) & tidally disrupted (10)$^\dagger$\\

Canes Venatici I & 216$\pm 8$  & 5.08$\pm 0.08$ & 647$\pm 27$ &  214 & (12) & (1), (11) & tidally disrupted (13)$^\dagger$\\

Coma Berenices   &  44$\pm 4$  & 3.46$\pm 0.24$ &  79$\pm  6$ &   59 & (12) & (1), (11) & tidally disrupted (12)$^\dagger$\\

Hercules         & 133$\pm 6$  & 4.42$\pm 0.16$ & 175$\pm 22$ &   30 & (12) & (1), (11) & \makecell{tidally disrupted (14)$^\dagger$ \\ high ellipticity}\\

Segue I          & 23$\pm  2$  & 2.54$\pm 0.32$ &  28$\pm  9$ &   71 & (15) & (1), (11) & high ellipticity\\

Tucana           & 887$\pm49$  & 5.75$\pm 0.08$ & 273$\pm 52$ &   62 & (16) & (1) & high ellipticity\\

Ursa Major I     & 94$\pm 4$   & 4.14$\pm 0.12$ & 190$\pm 46$ &   39 & (12) & (1), (11) & \makecell{tidally disrupted (17)$^\dagger$ \\ high ellipticity}\\

Willman I        & 38$\pm 7$   & 3.01$\pm 0.28$ &  25$\pm  5$ &   40 & (18) & (1) & \makecell{tidally disrupted (19)$^\dagger$\\ high ellipticity}\\ \addlinespace

NGC 147 & 712$^{+19}_{-21}$ & 7.84$\pm 0.04$ &  672$\pm  23$ & 520 & (20) & (20)\\

NGC 185 & 620$^{+18}_{-19}$ & 7.84$\pm 0.04$ &  565$\pm   7$ & 442 & (20) & (20)\\

NGC 205 & 824$\pm 27$       & 8.52$\pm 0.04$ &  594$\pm 107$ & 725 & (21) & (21)\\

And I   & 727$^{+17}_{-18}$ & 6.58$\pm 0.04$ &  772$\pm  85$ &  51 & (22) & (1) & tidally disrupted (23)$^\dagger$\\

And II  & 630$\pm 15$       & 6.85$\pm 0.08$ & 1355$\pm 142$ & 531 & (24) & (24)\\

And III & 723$^{+24}_{-18}$ & 5.89$\pm 0.12$ &  427$\pm  43$ &  62 & (22) & (1) & \makecell{tidally disrupted (23)$^\dagger$\\ high ellipticity}\\

And V   & 742$^{+22}_{-21}$ & 5.55$\pm 0.08$ &  365$\pm  57$ &  85 & (22) & (1) & tidally disrupted (23)$^\dagger$\\

And VI  & 783$\pm 25$       & 6.44$\pm 0.08$ &  537$\pm  54$ &  36 & (25) & (1) \\

And VII & 762$\pm 35$       & 6.98$\pm 0.12$ &  965$\pm  52$ & 136 & (22) & (1) \\

And IX  & 600$^{+23}_{-91}$ & 4.97$\pm 0.44$ &  582$\pm  23$ &  32 & (22) & (1) \\

And XIV & 793$^{+179}_{-23}$& 5.37$\pm 0.20$ &  434$\pm 212$ &  48 & (22) & (1) \\

And XXI & 827$^{+25}_{-23}$ & 5.85$\pm 0.24$ & 1004$\pm 123$ &  38 & (22), (25) & (1) \\

And XXIII & 748$^{+21}_{-31}$ & 6.00$\pm 0.20$ & 1034$\pm 97$ &  44 & (25) & (1) \\

Cetus   & 779$\pm 43$         & 6.44$\pm 0.08$ &  791$\pm 75$ & 116 & (26) & (1) \\
\bottomrule
\end{tabular}
\tabnote{Distance, V-band luminosity, and half-light radius are reported in \citet{2017ApJ...836..152L}. We note that we regard Cetus and Tucana dwarf spheroidal galaxies as "satellites" even though they are relatively isolated \citep{2017ApJ...836..152L}. For the first 8 classical dwarf spheroidal galaxies, $N_{\star}$ is the sum of the probabilities of the member stars. For the others, we count the member stars from the catalog. As And XXI is measured independently in two groups, we match the catalogs and use the combined data. dwarf spheroidal galaxies with high ellipticity are reported in \citet{2017ApJ...836..152L}. The $^\dagger$ represents the references reporting (tentative)evidence of tidal disruption.
\textbf{References:} (1) This work; (2) \citet{2009AJ....137.3100W}; (3) \citet{2009ApJ...704.1274W}; (4) \citet{2015MNRAS.448.2717W}; (5) \citet{2008ApJ...675..201M}; (6) \citet{2017ApJ...836..202S}; (7) \citet{2018AJ....156..257S}; (8) \citet{2023MNRAS.525.2875S}$^\dagger$; (9) \citet{2011ApJ...736..146K}; (10) \citet{2016MNRAS.461.3702R}$^\dagger$; (11) \citet{2015ApJ...801...74G}; (12) \citet{2007ApJ...670..313S}; (13) \citet{2012ApJ...744...96O}$^\dagger$; (14) \citet{2015ApJ...804..134R}$^\dagger$; (15) \citet{2011ApJ...733...46S}; (16) \citet{taibi2020tucana}; (17) \cite{2008A&A...487..103O}$^\dagger$; (18) \citet{2011AJ....142..128W}; (19) \citet{2006astro.ph..3486W}$^\dagger$; (20) \citet{2010ApJ...711..361G}; (21) \citet{2006AJ....131..332G}; (22) \citet{2012ApJ...752...45T}; (23) \citet{2006MNRAS.365.1263M}$^\dagger$; (24) \citet{2012ApJ...758..124H}; (25) \citet{2013ApJ...768..172C}; (26) \citet{2014MNRAS.439.1015K}}
\end{table*}

\begin{table}[t!]
\caption{Velocity dispersion profiles for the 18 galaxiesx in Figure \ref{fig:profiles}\label{tab:table2}}
\centering
\begin{tabular}{lrr}
\toprule
Name        & r (pc) & $\boldsymbol{\sigma_{\rm{los}}}$ (km s$^{-1}$) \\
\midrule
And I           &  151.9  & 11.56 $\pm$ 4.39      \\
And I           &  293.3  &  7.57 $\pm$ 2.56      \\
And I           &  420.4  &  3.90 $\pm$ 1.88      \\ 
And I           &  547.1  &  6.55 $\pm$ 1.94      \\
And I           &  687.6  &  6.15 $\pm$ 2.55      \\
And I           &  788.9  & 11.92 $\pm$ 3.35      \\
And I           &  929.7  & 16.68 $\pm$ 4.60      \\
And III         &  78.2   & 11.53 $\pm$ 2.94      \\
And III         &  140.7  &  8.36 $\pm$ 2.39      \\
...             &  ...    & ...                   \\
\bottomrule
\end{tabular}
\tabnote{
Only a portion of this table is shown here. A machine-readable version of the full table is available.
}
\end{table}

\subsection{DETERMINATION OF OBSERVED GRAVITATIONAL ACCELERATION\label{sec:observation}}

We now use the velocity dispersion profiles to derive the gravitational acceleration at each radial bin of dwarf spheroidal galaxies through dynamical mass analysis. To estimate the gravitational acceleration, we follow the approach of \citet{2010MNRAS.406.1220W}, which is based on the Jeans equations under the assumptions of spherical symmetry and dynamical equilibrium. In particular, we use the equation:
\begin{equation}
\label{eq:mass}
M\left(r\right)=\frac{r\sigma^2_r}{G}\left(\gamma_*+\gamma_\sigma-2\beta\right),
\end{equation}
where $\beta$ is the velocity anisotropy parameter defined by radial ($\sigma_r$) and tangential ($\sigma_t$) velocity dispersion as $\beta\left(r\right) \equiv 1 - \sigma^2_{t}/\sigma^2_{r}$. The $\gamma_* \equiv -\mathrm{d} \ln n_*/\mathrm{d} \ln r$ and $\gamma_\sigma \equiv -\mathrm{d} \ln \sigma^2_r/\mathrm{d} \ln r$ represent the slope of the stellar number density and radial velocity dispersion profile in the logarithmic scale, respectively. Equation~(\ref{eq:mass}) can be rearranged with the total velocity dispersion, $\sigma^2_\mathrm{tot} = \sigma^2_r + \sigma^2_\theta + \sigma^2_\phi = \left( 3 - 2\beta\right)\sigma^2_r$, which includes the $\beta$ dependence in its definition:
\begin{equation}
\label{eq:GM/r}
GM\left(r\right)r^{-1} = \sigma^2_\mathrm{tot}\left(r\right) + \sigma^2_r\left(r\right)\left(\gamma_\sigma + \gamma_* -3 \right).
\end{equation}
One can express the relation of mass estimates between the cases of  $\beta = 0$ and $\beta \ne 0$ as follows (see equation (20) of \citealt{2010MNRAS.406.1220W}):
\begin{equation} 
\label{eq:mass2}
lM\left(r;\beta\right)-M\left(r;0\right)=\frac{\beta\left(r\right)r\sigma^2_r}{G}\left(\gamma_*+\gamma_\sigma+\gamma_\beta-3\right)
\end{equation}
where $\gamma_\beta \equiv -\mathrm{d} \ln \beta/\mathrm{d} \ln r$. \cite{2010MNRAS.406.1220W} suggest that there could exist the characteristic radius, $r_{\rm{eq}}$, where $\left(\gamma_*+\gamma_\sigma+\gamma_\beta-3\right)$ goes to zero. Thus, the enclosed mass is insensitive to ${\beta}$ and the value of $r_{\rm{eq}}$ is close to the deprojected half-light radius, $r_{1/2}$.
Then the mass can be expressed only with the observed line-of-sight velocity dispersion, $\sigma_{\rm{los}}$\footnote{The total velocity dispersion, $\sigma_\mathrm{tot}$, can be expressed with line-of-sight velocity dispersion, $\sigma_\mathrm{los}$; see Sec.2.1 and 3.2 in \citet{2010MNRAS.406.1220W}.}, once we adopt the luminosity-weighted average on equation~(\ref{eq:GM/r}) together with scalar virial theorem to exploit the condition $\left<\left(\gamma_*+\gamma_\sigma-3\right)\sigma^2_\mathrm{r}\right> = 0$ (see \citealt{2010MNRAS.406.1220W} for more details).
\begin{equation}
\label{eq:Masswolf}
M\left(r_{1/2}\right) \simeq \frac{r_{1/2}\left<\sigma^2_\mathrm{tot}\right>}{G} \simeq \frac{3r_{1/2}\left<\sigma^2_\mathrm{los}\right>}{G}.
\end{equation}
%
We would like to use equation~(\ref{eq:Masswolf}) for any radius by assuming an isotropic velocity dispersion profile (i.e. $\beta(r)=0$) to make our computation of acceleration simple. 
Actually there are several studies indicating that ${\beta}$ is close to zero and the model with constant ${\beta}$ is consistent with observations \citep{2009MNRAS.394L.102L}; this includes the observations of dwarf spheroidal galaxies in \citet[Carina, Fornax, Leo I, Sculptor, and Sextan]{2015arXiv150408273M} and in \citet[Draco]{2024ApJ...970....1V}.
{We can then estimate the gravitational acceleration from the observed velocity dispersion using a simple form of the mass estimator without $\beta$:
\begin{equation}
\label{eq:gobs1}
\textsl{\textrm{g}}_\mathrm{obs}\left(r\right) = \frac{GM\left(r\right)}{r^2} = \frac{\sigma^2_\mathrm{tot}}{r} =\frac{3\sigma^2_\mathrm{los}}{r}.
\end{equation}

Here, the third term of $\boldsymbol{\sigma_\mathrm{tot}^2}$ is from the luminosity-weighted average on equation~(\ref{eq:GM/r}) and the scalar virial theorem, and the fourth term of $\boldsymbol{\sigma_\mathrm{los}^2}$ is from the assumption of constant $\boldsymbol{\beta}$ and spherical symmetry. We note that the observed $\sigma_\mathrm{los}$ profiles are projected ones, and the projected 2D radius $R$ can be replaced by the deprojected 3D radius $r$ under the assumption of spherical symmetry. As mentioned above, three galaxies in our sample (NGC 147, NGC 185, and NGC 205) are known to have non-negligible rotation \citep{2006AJ....131..332G,2010ApJ...711..361G}. To take into account this rotational contribution to the gravitational acceleration, we use the following equation for such systems:
\begin{equation}
\label{eq:gobs2}
\textsl{\textrm{g}}_\mathrm{obs} = \frac{3\sigma^2_\mathrm{los}}{r} + \frac{V^2_\mathrm{rot}}{r}.
\end{equation}
%
\begin{figure*}[ht!]
\centering
\includegraphics[width=165mm]{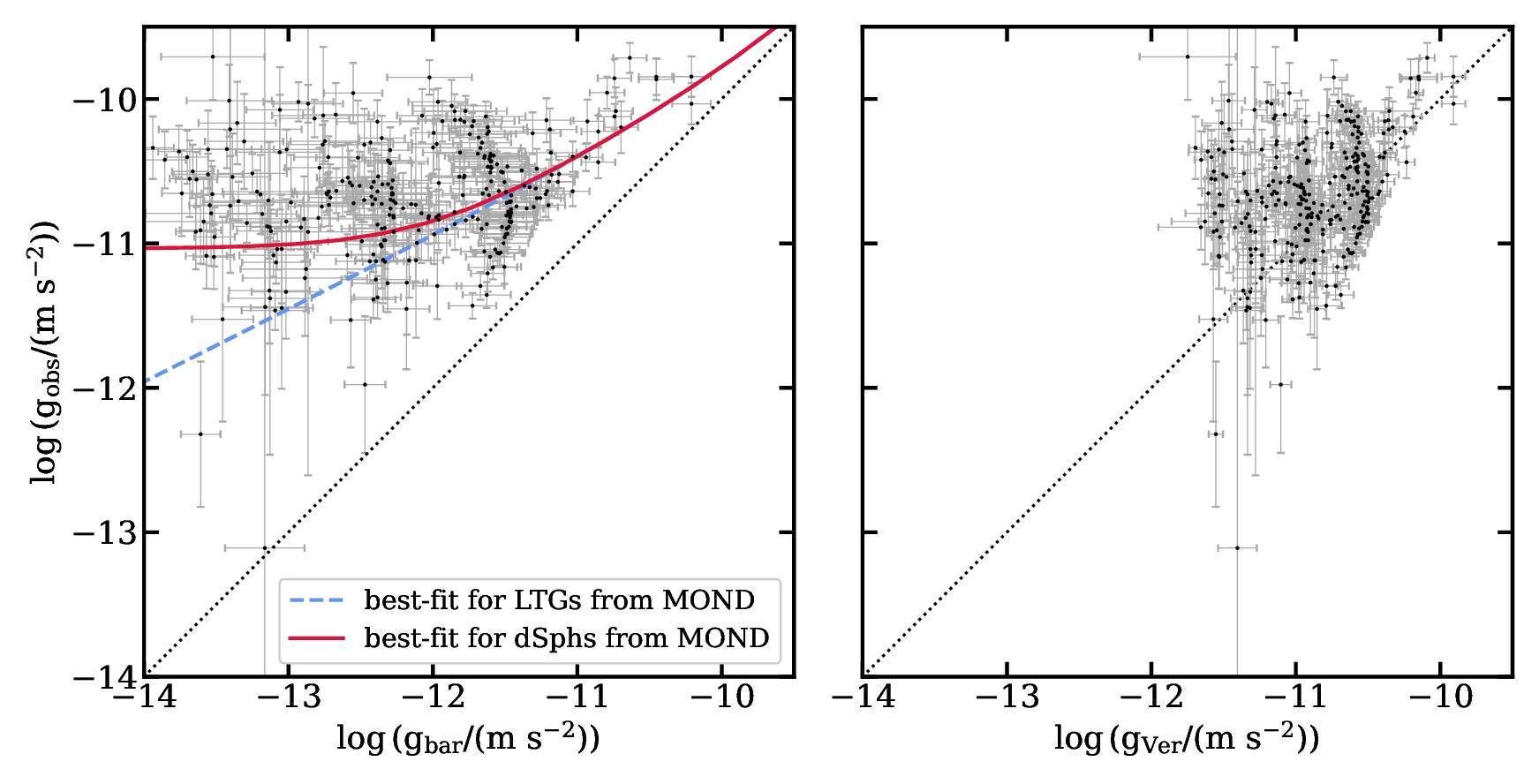}
\caption{The figure illustrates the radial acceleration relation between the predicted gravitational acceleration and the measured gravitational acceleration for 30 dwarf spheroidal galaxies. The prediction of the gravitational acceleration in the Newtonian framework is shown in the left panel. The blue dashed line represents the empirical best-fit for late-type galaxies, which is given by Equation~(\ref{eq:LTGfit}), while the red solid line shows the empirical best-fit for dwarf spheroidal galaxies, which is given by Equation~(\ref{eq:DSPHfit}). We note that Equation~(\ref{eq:DSPHfit}) contains two free parameters, namely $\textsl{\textrm{g}}_\dagger$ and $\hat{\textsl{\textrm{g}}}$, with the second parameter imposing the flattening. On the right panel, the predicted acceleration is shown by using Equation~\ref{eq:gver} based on the Verlinde framework. The first and the last points of the predicted acceleration for each galaxy are omitted by numerical limitations in evaluating $\textsl{\textrm{g}}_\textrm{Ver}$. Similarly, the same points in the observed acceleration are also omitted for consistency.\label{fig:fig2}}
\vspace{5mm} 
\end{figure*}
\subsection{PREDICTION OF THE GRAVITATIONAL ACCELERATION VIA BARYONIC COMPONENTS\label{subsec:prediction}} 

According to \citet{2017ApJ...836..152L}, the gravitational acceleration due to the baryonic (i.e., visible) matter of dwarf spheroidal galaxies within the half-light radius $r_{1/2}$ can be estimated as:
\begin{equation}
\label{eq:gbar1}
\textsl{\textrm{g}}_\mathrm{bar}\left(r_{1/2}\right) = -\nabla \Phi \left(r_{1/2}\right) = G\frac{\Upsilon_V L_V}{2r^2_{1/2}}
\end{equation}
where $\Upsilon_V$ is the V-band stellar mass-to-light ratio, and $L_V$ is the V-band luminosity; the resulting mass of the baryonic component becomes $M_\mathrm{bar} = \Upsilon_V L_V$. Here, we assume a mass-to-light ratio of $\Upsilon_V = 2.0\pm0.5\ M_\odot/L_\odot$ for all 30 dwarf spheroidal galaxies \citep{2017ApJ...836..152L}.

We further extend this relation to derive the radial profile of baryonic mass from that of stellar light. We further assume that the stellar density profile of dwarf spheroidal galaxies in the Local Group follows a spherically symmetric density profile of Plummer \citep{1911MNRAS..71..460P}, which has been commonly used for dwarf spheroidal galaxies \citep{1997MNRAS.286..669G,1999PASJ...51..943S,2011MNRAS.413.2606A}. The mass profile can be obtained by integrating the density profile:
\begin{equation}
\label{eq:Mbar}
M_\mathrm{bar}\left(r\right) = \frac{r^3 \Upsilon_V L_V}{\left(r^2 + r_\mathrm{{half}}^2\right)^{3/2}}.
\end{equation}
We calculate the baryonic acceleration:
\begin{equation}
\label{eq:gbar2}
\textsl{\textrm{g}}_\mathrm{bar}\left(r\right) = \frac{GM_\mathrm{bar}\left(r\right)}{r^2}.
\end{equation}
When we consider the gravitational field in the Verlinde framework, we need to have an additional term to this equation, which is called the apparent dark matter. \citet{2024PDU....4501551Y} proposes a formulation for the gravitational field in the emergent gravity as: 
\begin{equation}
\label{eq:gver}
\textsl{\textrm{g}}_\mathrm{Ver}^2=\textsl{\textrm{g}}_\mathrm{bar}^2 + \textsl{\textrm{g}}_\mathrm{ad}^2  
\end{equation}
where $\textsl{\textrm{g}}_\mathrm{bar}$ is the gravitational acceleration from baryonic matter in the Newtonian framework as shown above, and $\textsl{\textrm{g}}_\mathrm{ad}$ comes from apparent dark matter, as first suggested in \citet{10.21468/SciPostPhys.2.3.016}. The above equation is derived by expanding the derivation of $\textsl{\textrm{g}}_\mathrm{Ver}$ in \citet{10.21468/SciPostPhys.2.3.016}, which used the theory of elastic inclusions. We follow the method in \citet{2023CQGra..40bLT01Y} to compute the gravitational acceleration predicted from the emergent gravity. Specifically, we use the following equation to derive $\textsl{\textrm{g}}_\mathrm{ad}$:
\begin{eqnarray}
\label{eq:gad}
\textsl{\textrm{g}}_\mathrm{ad}^2 = \frac{a_0}{6} \left( 2 \textsl{\textrm{g}}_\mathrm{bar} - \partial_r\Phi_\mathrm{bar} \right), \nonumber\\ 
\Phi_\mathrm{bar} = - \frac{2\textsl{\textrm{g}}_\mathrm{bar}^2}{4 \pi G \rho_\mathrm{bar} - \partial_{r} \textsl{\textrm{g}}_\mathrm{bar}}
\end{eqnarray}
where $a_0$ is the characteristic acceleration scale suggested by Verlinde. Note that the non-linearity manifests in the above equations. As in MOND (see Equation (\ref{eq:LTGfit})), this violation of the superposition principle is crucial to explain the Tully-Fisher relation. Also, $\Phi_\mathrm{bar}$, which is related to the space-space component of metric, must not be confused with the Newtonian potential $\Phi$, related to the time-time component of metric. In the presence of spherical symmetry, the former reduces to $\Phi_\mathrm{bar}=-GM_\mathrm{bar}\left(r\right)/r$, while the latter reduces to $\Phi=\int dr\,GM_\mathrm{bar}\left(r\right)/r^2$.

\begin{figure*}[t!]
\centering
\includegraphics[width=175mm]{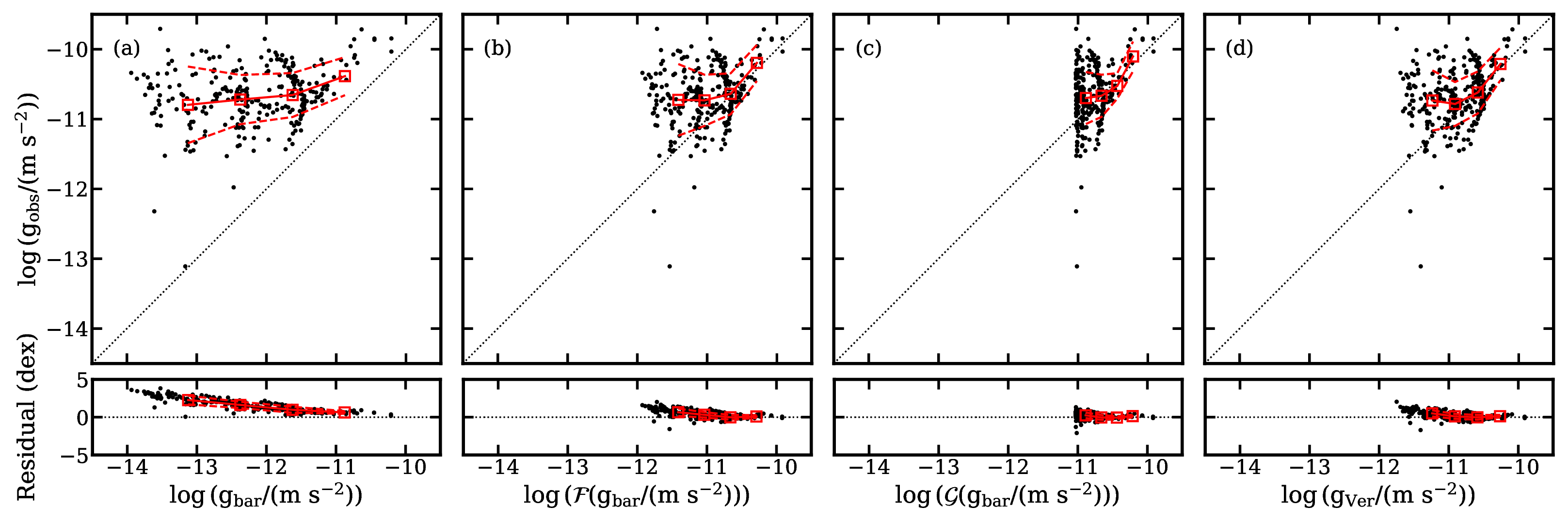}
\caption{The top panels display a plot similar to Figure~\ref{fig:fig2}. The red squares represent the median for each bin, and the red dashed lines indicate the 1$\sigma$ deviation. Each panel shows different predictions; (a): Newtonian gravity, (b): MOND interpolating function for late-type galaxies, (c): MOND interpolating function for dwarf spheroidal galaxies, (d): Emergent gravity. The bottom panel represents the observational offset from the prediction. The residuals in each case are much larger than those for spiral galaxies (see \citealt{2023CQGra..40bLT01Y}).\label{fig:fig3}}
\vspace{5mm} 
\end{figure*}

For more accurate calculation of the gravitational acceleration from the emergent gravity, we also consider the quasi-de Sitter universe as in \citet{2023CQGra..40bLT01Y} and \citet{2018MNRAS.477.1285D}. The quasi-de Sitter universe is based on the assumptions that the positive cosmological constant represents dark energy, and that cosmic evolution explained by general relativity adequately describes the quasi-de Sitter universe. Here the quasi-de Sitter value of $a_0$, $a_0 = 5.41 \pm 0.06 \times 10^{-10} \mathrm{m\ s^{-2}}$, is obtained by assuming that only the cosmological constant part contributes to $a_0$, which is estimated from Planck observations \citep{2018MNRAS.477.1285D}. However, the precise value of $a_0$ remains unknown, as the relativistic version of Verlinde's emergent gravity, which is necessary to apply it to the universe as a whole in the FRW framework, has not been formulated yet. At this point, it is also unclear whether $a_0$ depends on time (i.e., redshift).

\section{RESULTS\label{sec:result}} 

We obtain the gravitational acceleration at different galactocentric radii from the velocity dispersion profiles of 30 galaxies, of which we include the contribution from rotation for three dwarf spheroidal galaxies. Figure~\ref{fig:fig2} shows a relation between the observed gravitational acceleration and the prediction based on Newtonian gravity and emergent gravity, respectively, which is commonly referred to as the radial acceleration relation \citep{2016PhRvL.117t1101M,2017ApJ...836..152L,2022NatAs...6...35L}. The left panel of Figure~\ref{fig:fig2} shows the observed acceleration at different galactic radii as a function of the expected acceleration from baryonic matter only.  To obtain the error associated with the acceleration estimate, we use a Python package \texttt{Uncertainties}\footnote{\url{http://pythonhosted.org/uncertainties/}} to account for the errors in the velocity dispersion estimates as well as other input data (e.g., half-light radius, luminosity, and so on). It should be noted that the inner and outer most points of the radial profile are excluded from the calculation of gravitational acceleration because of numerical problem in deriving the radial derivatives.

\begin{figure*}[ht!]
\centering
\includegraphics[width=165mm]{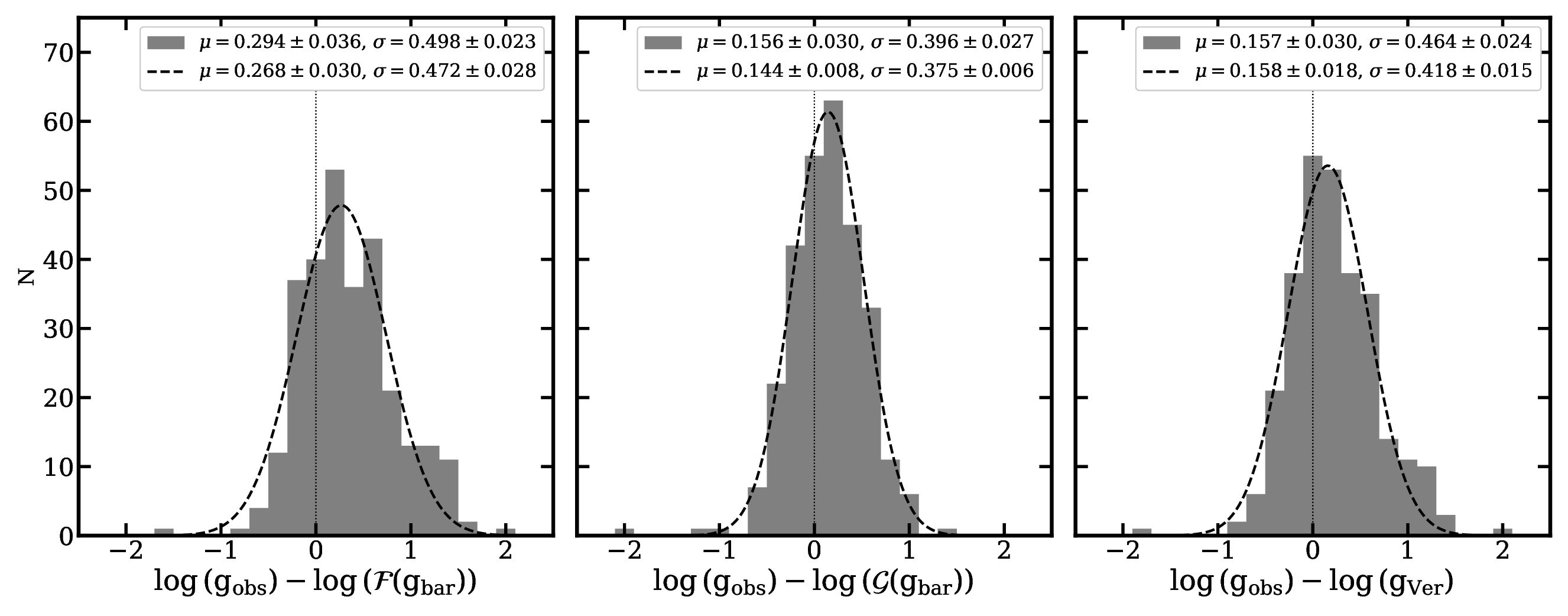}
\caption{In the left and middle panels, we present histograms showing the difference between the observed acceleration $\textsl{\textrm{g}}_\textrm{obs}$ and the two MOND interpolating functions for dwarf spheroidal galaxies. The left panel uses Equation~(\ref{eq:LTGfit}) originally obtained from late-type galaxies, and the middle panel uses Equation~(\ref{eq:DSPHfit}) obtained from dwarf spheroidal galaxies. In the right panel, we show a similar histogram, but for the difference between $\textsl{\textrm{g}}_\textrm{obs}$ and $\textsl{\textrm{g}}_\textrm{Ver}$. We evaluate the mean and standard deviation of the histogram using the bootstrap resampling method. The dashed line in each panel represents the Gaussian distribution corresponding to the histogram.\label{fig:fig4}}
\vspace{5mm} 
\end{figure*}

In the left panel of Figure~\ref{fig:fig2}, we also show the best-fit lines of MOND from \citet{2017ApJ...836..152L}: blue dashed and red solid lines for late-type galaxies and dwarf spheroidal galaxies, respectively. An interpolating function for late-type galaxies is given by,
\begin{equation}
\label{eq:LTGfit}
\textsl{\textrm{g}}_\mathrm{obs} = \mathcal{F}\left(\textsl{\textrm{g}}_\mathrm{bar}\right) = \frac{\textsl{\textrm{g}}_\mathrm{bar}}{1 - e^{-\sqrt{\textsl{\textrm{g}}_\mathrm{bar}/\textsl{\textrm{g}}_\dagger}}}
\end{equation}
where $\textsl{\textrm{g}}_\dagger$ is a free parameter and has a value of $1.20 \pm 0.02 \times 10^{-10}\ \mathrm{m\ s^{-2}}$. On the other hand, the interpolating function for dwarf spheroidal galaxies is given by, 
\begin{equation}
\label{eq:DSPHfit}
\mathcal{G}\left(\textsl{\textrm{g}}_\mathrm{bar}\right) = \frac{\textsl{\textrm{g}}_\mathrm{bar}}{1 - e^{-\sqrt{\textsl{\textrm{g}}_\mathrm{bar}/\textsl{\textrm{g}}_\dagger}}} + \hat{\textsl{\textrm{g}}}\ e^{-\sqrt{\textsl{\textrm{g}}_\mathrm{bar}\textsl{\textrm{g}}_\dagger/\hat{\textsl{\textrm{g}}}^2}}
\end{equation}
where $\textsl{\textrm{g}}_\dagger$ and $\hat{\textsl{\textrm{g}}}$ are two free parameters with values of $1.1 \pm 0.1 \times 10^{-10}\ \mathrm{m\ s^{-2}}$ and $9.2 \pm 0.2 \times 10^{-12}\ \mathrm{m\ s^{-2}}$, respectively. 

As expected, the data follow the best-fit line for dwarf spheroidal galaxies (i.e. red solid line) at $\textsl{\textrm{g}}_\textrm{bar} < 10^{-10}\ \mathrm{m\ s^{-2}}$ better than the line for late-type galaxies(i.e. blue dashed line). \citet{2017ApJ...836..152L} note that the cause of the flattening observed in the region for low acceleration is not clearly known, but may be because of the interaction of DM halos in the $\Lambda$CDM context or intrinsic limitations in the data. The right panel of Figure~\ref{fig:fig2} shows the relation similar to the left panel, but for the expectation from the emergent gravity. Despite the large scatter, the data move toward the one-to-one relation compared to those in the left panel. We also compute the Pearson correlation coefficient to quantify the correlation. The resulting coefficient is 0.30 with the p-value of <0.001, which suggests a weak correlation.

The top panels of Figure~\ref{fig:fig3} show the plot similar to Figure~\ref{fig:fig2}, but focus more on the trend of the data without error bars for clarity. The bottom panels show the difference between predictions and the observations. The residuals in the left-most panel (i.e. Newtonian gravity) of Figure~\ref{fig:fig3} are significantly larger than those in the right-most panel (i.e. emergent gravity). This feature is similar to the late-type galaxies in \citet{2023CQGra..40bLT01Y}. We note that the interpolating function for dwarf spheroidal galaxies is truncated at the scale $\rm {log}\left(\mathcal{G}\left(\textsl{\textrm{g}}_\mathrm{bar}\right)\right) \sim -11$, which is the impact of the second term in Equation~(\ref{eq:DSPHfit}). The panels (b), (c), and (d) show an offset between predictions and observations on the scale $\lesssim -11$.

To better illustrate the difference between observation and expectation with different models, we show the histograms of the residuals along with the best-fit Gaussian curves in Figure~\ref{fig:fig4}. The plot shows that the histogram in each panel follows well the dashed line representing the best-fit Gaussian distribution\footnote{obtained using the curve-fit module in scipy.optimize}. To evaluate the confidence in the predicted acceleration in each framework, we use the bootstrap resampling method to estimate the error of the median and the variance of the residuals. The median of the residuals in the central and right panels are nearly zero, contrasting with the non-zero median in the left panel. Moreover, the variance in the central and right panels is less than that in the left panel. As noted before, this indicates a better fit for the right panel compared to the left one, suggesting that the interpolating function derived from late-type galaxies is less effective than $\textsl{\textrm{g}}_\textrm{Ver}$ in dwarf spheroidal galaxies.

To further improve the correlation of gravitational accelerations between observations and predictions from the emergent gravity (i.e. right-most panel of Figure \ref{fig:fig3}), we make a subsample by removing some galaxies that can potentially contaminate the correlation. For example, many dwarf galaxies are known to experience tidal disturbance because of their host galaxies (e.g. Hercules dwarf galaxy associated with the Milky Way - \cite{2015ApJ...804..134R}); the observed velocity dispersion of these galaxies may not be translated as gravitation acceleration. Among the 30 galaxies in our sample, there are 10 galaxies that might have experienced such tidal interactions (see \textbf{Notes} column in Table~\ref{tab:table1}). We also select 7 galaxies that have high ellipticity in \citet{2017ApJ...836..152L}, which could be again gravitationally affected by neighboring galaxies.

\begin{figure*}[t!]
\centering
\includegraphics[width=165mm]{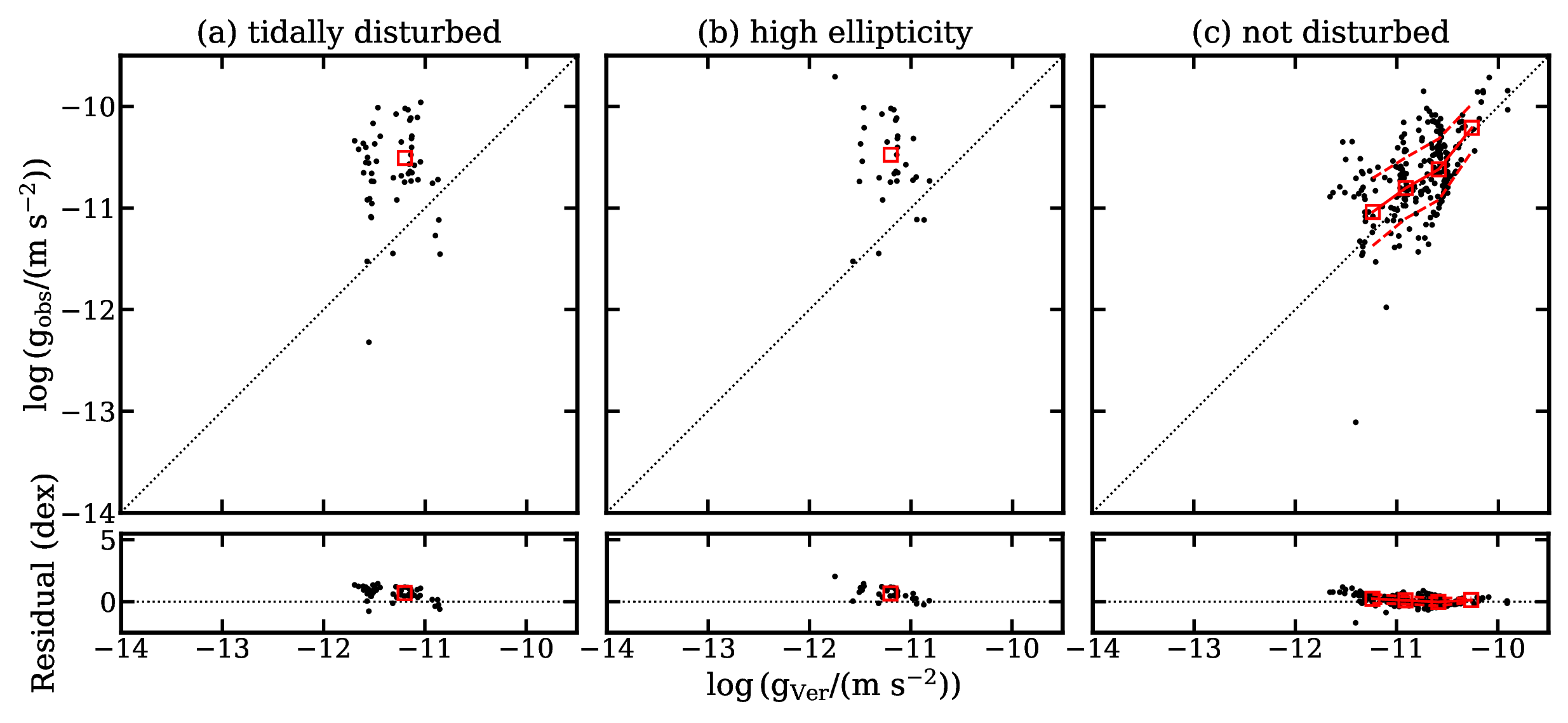}
\caption{Similar to Figure~\ref{fig:fig3}.  (a): tidally disturbed dwarf spheroidal galaxies reported in the literature.  (b): dwarf spheroidal galaxies with high ellipticity reported in \citet{2017ApJ...836..152L}.  (c): dwarf spheroidal galaxies without tidal disturbance. \label{fig:fig5}}
\vspace{5mm} 
\end{figure*}
\begin{figure*}[t!]
\centering
\includegraphics[width=165mm]{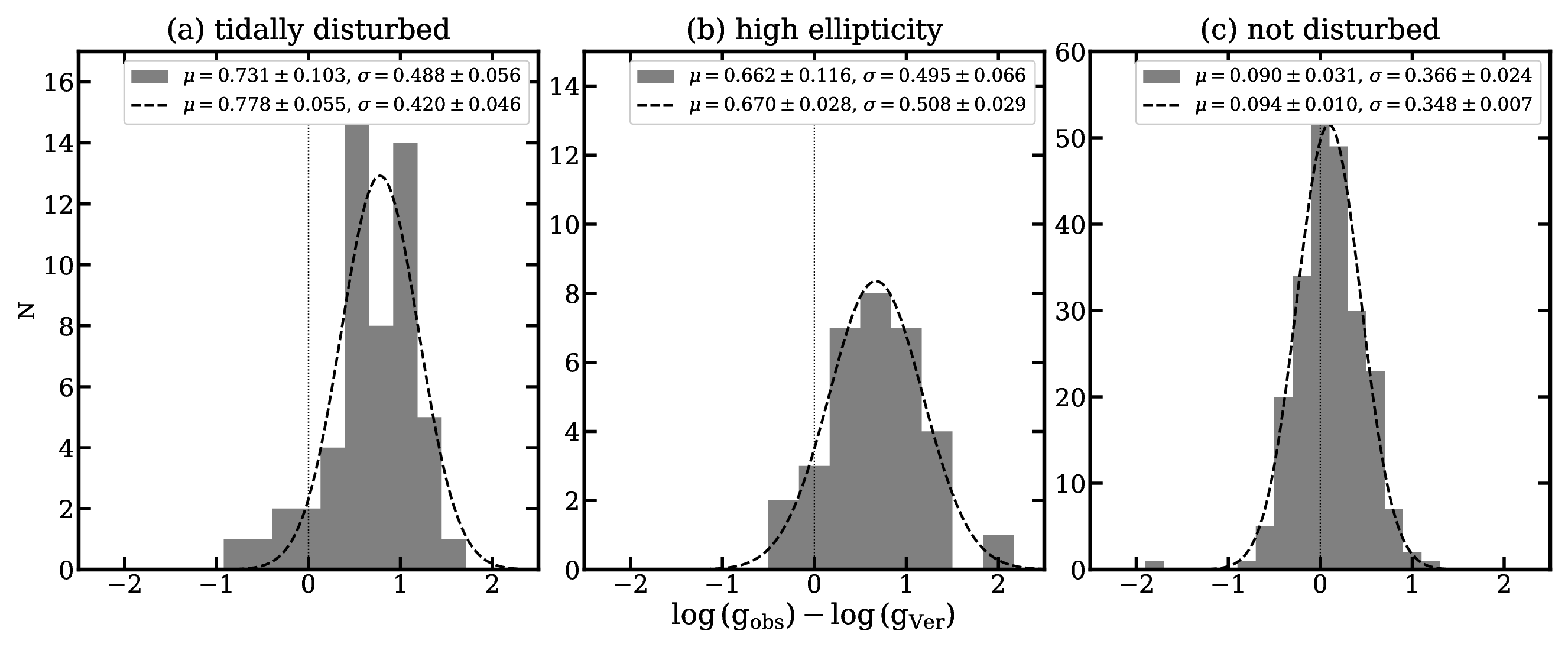}
\caption{Similar to Figure~\ref{fig:fig4}. Each panel corresponds to the ones described in Figure~\ref{fig:fig5}. \label{fig:fig6}}
\vspace{5mm} 
\end{figure*}

Figures~\ref{fig:fig5} and \ref{fig:fig6} show three subsets, respectively: (a) shows dwarf spheroidal galaxies undergoing the tidal disturbance by the host galaxy, (b) indicates dwarf spheroidal galaxies with high ellipticity, and (c) illustrates the remaining dwarf spheroidal galaxies without the effect, of or less sensitive to, the tidal disturbance and with low ellipticity. When we compare it with the right-most panel of Figure~\ref{fig:fig3}, we find that the offset is reduced, hence the overall feature is closer to an one-to-one relation. The subset (c) shows a much stronger correlation between prediction and observation. The Pearson correlation coefficient is 0.51 with a p-value of <0.001, indicating a better correlation than the case of all the galaxies including those with tidally disturbance and high ellipticity.
 
\section{DISCUSSION\label{discussion}}

We investigate whether the emergent gravity can explain the observed gravitational acceleration even in the scale of less massive system (i.e. dwarf galaxies), and find the positive results. However, our results are based on several assumptions, and need to be examined with more data. Here, we discuss the caveats of our analysis.

\begin{figure}[t!]
\centering
\includegraphics[width=80mm]{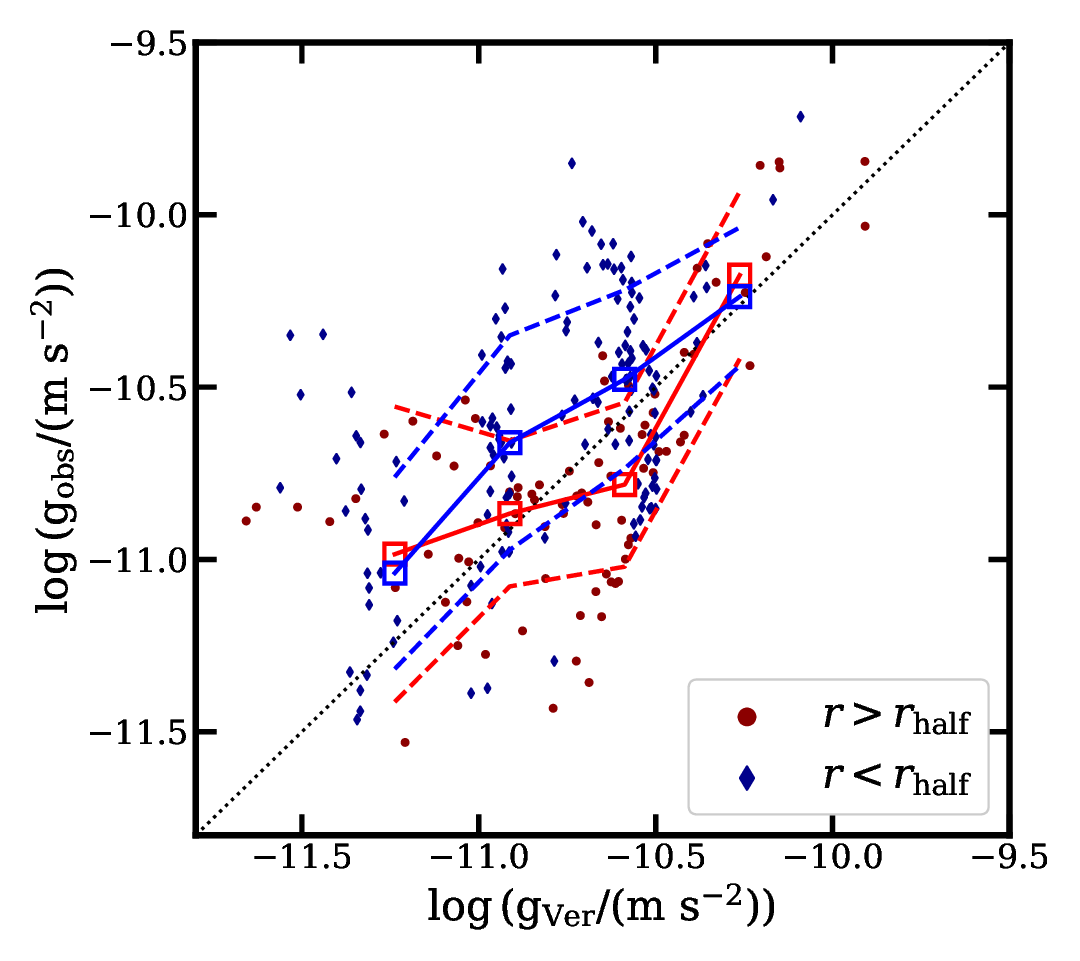}
\caption{We divide the region by 3D half-light radius of each dwarf spheroidal galaxies, whose value is reported in Table~\ref{tab:table1}. We use only the galaxies that are not tidally disturbed. The overall trend is consistent with each other in the 1$\sigma$ range.\label{fig:fig7}}
\vspace{5mm} 
\end{figure}

Firstly, we use the radial profiles of the velocity dispersion of dwarf spheroidal galaxies. \citet{2009ApJ...704.1274W} provide high-quality data set of velocity dispersion profiles for eight dwarf spheroidal galaxies. \citet{2006AJ....131..332G}, \citet{2010ApJ...711..361G} and \citet{2012ApJ...758..124H} also provide high-quality data for four dwarf spheroidal galaxies of M31. Furthermore, we estimate other dwarf spheroidal galaxies by adopting the maximum likelihood estimation process as \citet{2006AJ....131.2114W} did. However, there are only 13 dwarf spheroidal galaxies that have member stars more than 150; the relatively smaller number of member stars for the remaining 17 galaxies can cause larger errors in velocity dispersion estimates (eventually larger scatter for the observed gravitational acceleration). 

We assume that the mass-to-light ratio remains constant for all dwarf galaxies, with a value of $2.0\pm0.5\ M_\odot/L_\odot$. Furthermore, we also assume that the mass-to-light ratio remains constant regardless of the radius. It is acknowledged that these assumptions could potentially lead to a significant degree of scatter in the radial acceleration relation.

We assume that dwarf spheroidal galaxies are in dynamical equilibrium, and have isotropic velocity anisotropies to compute the gravitational accelerations with Equations~(\ref{eq:gobs1}) and (\ref{eq:gobs2}). However, these assumptions may not be valid because of various interactions with neighboring galaxies (e.g., the host galaxy); this can introduce a bias in the estimate of gravitational acceleration. Indeed, some studies reported dwarf galaxies with tidal features \citep{2007ApJ...670..313S,2008A&A...487..103O,2011ApJ...733...46S,2015ApJ...804..134R,2016MNRAS.461.3702R}. The tidal disturbance can make the velocity dispersion to be estimated higher than expected from the mass of a galaxy \citep{2017ApJ...836..152L}. As Figures~\ref{fig:fig5} and \ref{fig:fig6} show, the galaxies with tidal disturbance (i.e. left and middle panels) can have observed acceleration higher than the one expected from their masses. The systematic offset in the framework of emergent gravity could be partly contributed by the overestimation due to tidal interactions. If we exclude such galaxies with tidal disturbance, we find that the agreement of the gravitational acceleration between the emergent gravity and the observation is better (right panel). It is noted that we do not consider the "external field effect", which could in principle occur in the context of the emergent gravity \citep{2017PhRvD..95f4019H} and thus affect the internal dynamics of dwarf spheroidal galaxies. Samples that are tidally disturbed tend to also have stronger external field effect, which exacerbates the bias in the estimate of gravitational acceleration. Further work would be required to accompany the external field effect.

We also compare our computations at different radial ranges of galaxies in Figure~\ref{fig:fig7}. The observed accelerations between inner and outer regions are consistent considering the 1$\sigma$ scatter even though there is a hint that the observed acceleration of the inner (i.e. blue diamonds and blue lines) region appears systematically higher than that for the outer region (i.e. red dots and red lines). We find a difference between our results and the prediction proposed by \citet{2018MNRAS.477.1285D}. While both studies obtain similar baryonic mass estimates, the approach used to derive $\textsl{\textrm{g}}_\textrm{obs}$ is slightly different. In this study, $\textsl{\textrm{g}}_\textrm{obs}$ is obtained directly from the observational data, whereas \citet{2018MNRAS.477.1285D} use an interpolating function separated at the half-light radius. Furthermore, the radial coverage in this study is limited by the range where $\sigma_\mathrm{los}$ is measured, which affects the predicted acceleration range. This suggests that the emergent gravity works well regardless of the radial distance of galaxies.

Here, we consider the quasi-de Sitter universe within the framework proposed by Verlinde, as explored by both \citet{2018MNRAS.477.1285D} and \citet{2023CQGra..40bLT01Y}. This model is a more realistic representation of the universe than Verlinde's original dS universe, which only consists of dark energy, without any matter or radiation. Our results complement the recent emergent gravity explanation of the binary orbit anomaly by \cite{2024PDU....4501551Y}, which is on a much shorter scale but at a much stronger acceleration scale than the current study. Thus, the emergent gravity is tested in an acceleration range above 5 dex and in a distance range above 7 dex.

It is noteworthy that Equation~(\ref{eq:LTGfit}) provides an excellent explanation of the dynamics of late-type galaxies by incorporating the free parameter $\textsl{\textrm{g}}_\dagger$. However, this same equation does not yield a comparable level of accuracy in describing the dynamics of dwarf spheroidal galaxies. This indicates that an additional term is required, along with another free parameter, $\hat{\textsl{\textrm{g}}}$, in order to achieve a satisfactory fit to the observed data. Verlinde's emergent gravity does not allow for empirical adjustment, although it provides an explanation for the diverse types of galaxies. $a_0$, the only free parameter needed in the emergent gravity, is solely determined by the expansion of the universe, which has a priori no reason to be related to galaxy rotation curves. On the other hand, $\textsl{\textrm{g}}_\dagger$ and $\hat{\textsl{\textrm{g}}}$ are introduced only for the purpose of fitting the galaxy rotation curves and imposing the low-acceleration flattening.

To repeat, MOND, i.e., Equation (\ref{eq:LTGfit}) underestimates the gravitational accelerations of dwarf spheroidal galaxies, in which case the emergent gravity prediction is more accurate. Therefore, considering that the gravitational accelerations of dwarf spheroidal galaxies are tiny, one may be tempted to conclude that MOND underestimates gravitational accelerations in a very low acceleration regime. However, according to the weak lensing analysis of \cite{2021A&A...650A.113B}, MOND and Verlinde's emergent gravity both predict the gravitational accelerations more or less correctly even in a very low acceleration regime. We also remark that it is well-known that MOND also underestimates the gravitational accelerations in galaxy clusters \citep{2005MNRAS.364..654P,2003MNRAS.342..901S,2007MNRAS.380..331S,1988AJ.....95.1642T}. Interestingly enough, galaxy clusters, as well as dwarf spheroidal galaxies, are cases where the spherical symmetry is present for the distribution of matter. Thus, it is likely that MOND fails, but Verlinde’s emergent gravity succeeds in the presence of spherical symmetry. We confirmed this for spheroidal galaxies in this study, but not for galaxy clusters. Several studies have utilized Verlinde's emergent gravity in analyzing galaxy clusters \citep{2016arXiv161209582M,2019JCAP...05..053T,2019PhRvD.100j4049I,2024JCAP...10..030G}. However, it is uncommon to find studies that simultaneously apply both this theory and MOND to ascertain which fits more accurately with galaxy clusters. This area appears to be suitable for future investigation.

\section{CONCLUSIONS\label{conclusions}}

In this study, we investigate the validity of emergent gravity at the scale of dwarf galaxies. We utilize kinematic data of dwarf satellite galaxies in the Local Group and adopt the Plummer density profile for stellar components. Our results are as follows:\\

(i) We find that the radial acceleration relation of dwarf galaxies is explained by the emergent gravity. The kinematics of dwarf galaxies are well described by the emergent gravity with comparing the observations. \\

(ii) We make subsample by removing some galaxies whose dynamics can be disturbed by their host galaxies, find the improvement of the correlation between observations and predictions with the emergent gravity.\\

(iii) We point out that the emergent gravity may provide an explanation for the dynamics of both late-type and dwarf galaxies. However, it should be noted that, in order to apply MOND to these galaxies, an interpolating function must be used that is specific to each type (see Equations ~(\ref{eq:LTGfit}) and ~(\ref{eq:DSPHfit})). This implies that MOND may require the constraint when applied to different types of galaxies.\\

Our results support the validity of the emergent gravity on much smaller galactic scales. However, there remain many observations to be explained by the emergent gravity that requires more exploration; e.g. formation and evolution of large-scale structure, cosmic microwave background (CMB) anisotropy. This requires the formulation of the relativistic version of the emergent gravity.


\acknowledgments
We thank the referee for helpful comments that improved the manuscript. We also thank D. Simon for offering the stellar-kinematic data and the SNU ExGalCos team members for valuable comments. 
We acknowledge the support of the National Research Foundation of Korea (NRF) grant funded by the Korea government (MSIT), [NRF-2021R1A2C1094577(SH, HSH), NRF-2022R1A2C1092306(YY)], Samsung Electronic Co., Ltd. (Project Number IO220811-01945-01, HSH), and Hyunsong Educational \& Cultural Foundation (HSH).




\bibliography{biblography}





\end{document}